\documentclass{sig-alternate-10pt}

\pdfpagewidth=\paperwidth
\pdfpageheight=\paperheight

\usepackage{graphicx}
\usepackage{multirow}
\usepackage{color}
\usepackage{paralist}
\usepackage{algorithmic}
\usepackage{caption}
\usepackage{subcaption}
\usepackage{xspace}
\usepackage{booktabs}
\usepackage{listings}
\usepackage{hyperref}
\usepackage{listings}
\usepackage{enumitem}
\makeatletter
\def\and{%
  \end{tabular}%
  \hskip 1em \@plus.17fil\relax
  \begin{tabular}[t]{c}}
\makeatother

\definecolor{Ora}{cmyk}{0, 0.6, 0.8, 0}
\definecolor{mygray}{gray}{0.5}

\newcommand{\remove}[1]{}

\newcommand{\myitem}[1]{\vspace*{0.07in}\noindent\textbf{#1}}

\title{Supercharge me: Boost Router Convergence with SDN}

\author{
Michael Alan Chang\\
\affaddr{Princeton University/ETH Z\"urich}
\and
Thomas Holterbach\\
\affaddr{ETH Z\"urich}
\and
Markus Happe\\
\affaddr{ETH Z\"urich}
\and
Laurent Vanbever\\
\affaddr{ETH Z\"urich}
}

\begin{document}
\thispagestyle{empty}
	
\maketitle

\begin{abstract}
Software Defined Networking (SDN) is a promising approach for improving the performance and manageability of future network architectures. However, little work has gone into using SDN to improve the performance and manageability of existing networks without requiring a major overhaul of the existing network infrastructure.

In this paper, we show how we can dramatically improve, or \textit{supercharge}, the performance of existing IP routers by combining them with SDN-enabled equipment in a novel way. More particularly, our supercharged solution substantially reduces the convergence time of an IP router upon link or node failure without inducing any reconfiguration of the IP router itself. Our key insight is to use the SDN controller to precompute backup forwarding entries and immediately activate them upon failure, enabling almost immediate data-plane recovery, while letting the router converge at its typical slow pace. By boosting existing equipment's performance, we not only increase their lifetime but also provide new incentives for network operators to kickstart SDN deployment.

We implemented a fully functional ``supercharger'' and use it to boost the convergence performance of a Cisco Nexus 7k router. Using a FPGA-based traffic generator, we show that our supercharged router systematically converges within {\raise.17ex\hbox{$\scriptstyle\sim$}}150ms, a 900$\times$ reduction with respect to its normal convergence time under similar conditions.
\end{abstract}

\section{Introduction} 
\label{sec:introduction}


By enabling logically-centralized and direct control of a
network forwarding plane, Software-Defined Networking (SDN) holds great
promises in terms of improving network management and performance, while
lowering costs at the same time. Realizing this vision is challenging though as
SDN requires major changes to a network architecture before
the benefits can be realized~\cite{vissicchio2014opportunities}. This is problematic as existing networks tend to have a huge
installed base of devices, management tools, and human operators that are not
familiar with SDN, leading to significant deployment hurdles. As a result, the
number of SDN deployments has been rather limited in scope; there have been efforts in private
backbones~\cite{google_b4_sigcomm2013, microsoft_swan_sigcomm2013} and software
deployments at the network edge~\cite{Casado:2012:FRE:2342441.2342459}.

In order to kickstart a wide-scale SDN deployment, we argue that operators need
to be offered with SDN-based technologies possessing at least three key
characteristics. First, the advantages of SDN should be readily apparent with only a \emph{small deployment}. Ideally, benefits should be reaped with the deployment of a
single SDN device; as comfort and enthusiasm increases, new SDN devices can be incrementally deployed. Second, they should be \emph{low-risk}. In particular,
they should require minimum changes to existing operational practices and
should be compatible with currently deployed technologies. Finally, they
should offer a \emph{high return}, meaning the SDN-based technologies should
solve a timely problem.

As an example of such a technology, we show how we can significantly improve
the performance of existing IP routers, \emph{i.e.} ``supercharge'' them, by
combining them with SDN-enabled devices. Supercharging a router is a low-risk,
high-reward operation. First, it provides operators with a strong incentives to
deploy SDN-enabled device as they enable them to increase the lifetime of their
routers, at a considerably lower cost than buying new ones\footnote{Current
SDN switches are orders of magnitude cheaper than fully equipped routers.}.
Second, supercharging a router does not change the existing router's behavior, just its
performance. Consequently, network operators can conveniently troubleshoot and maintain the original network. Third, once enough routers have been supercharged, those deployed SDN equipments can be used to implement a more disruptive SDN architecture.

In this short paper, we supercharge one particular aspect of the router
performance: its convergence time after a link or a node failure. Current
routers are often slow to converge after a link failure because of the time it
takes to update their forwarding tables; this is an entry-by-entry process that can go on for potentially hundreds thousand of entries. Our key insight is that, by coupling together a
router and a SDN switch, we can build a 2-stage forwarding table which spans
across the two devices with a first lookup done in the router and the second
one in the switch. With this type of hierarchical FIB, one can speed up the convergence by
tagging entries with the same primary and backup Next-Hop (NH) in the first
table, and then actually direct the traffic to the primary or backup NH in the second table. This way, if the primary NH fails, only the few
entries on the switch have to be updated. One contribution of our work is to
show how we can provision those tagging entries in a router using only a vanilla
routing protocol.

Besides convergence, several other aspects of a router performance can be
``supercharged'' by having a 2-stage forwarding table. Among others, the size
of the router forwarding tables can be increased using a SDN switch as a cache
(similarly to~\cite{ballani2009making}). In this case, the router table would
contain aggregated entries that would get resolved in the switch table.
Similarly, poor load-balancing decisions made by routers due to sub-optimal
stateless hash-functions~\cite{rfc2992, cao2000performance} can be overwritten
dynamically as the traffic traverses the neighboring SDN switch leading to
better network utilization. In all three examples, the factor limiting the
performance is the \emph{hardware design} itself, \emph{i.e.}, the forwarding
table organization, its forwarding table size, or the hash function used by the router. Unlike
software, this cannot be improved without buying new equipment, hence the
interest.

In~\cite{sdx_sigcomm2014}, Gupta \emph{et al.} used a similar technique to
scale an SDN-based Internet Exchange Point with the aim of decreasing the amount of forwarding rules that has to be maintained in the SDN switch by leveraging neighboring router resources. While we target convergence, not space, our contribution is also the opposite of theirs. We show how a SDN switch can improve the performance of the router. As such, our work nicely complements theirs.

\begin{figure}[t]
 \centering
 \includegraphics[width=\columnwidth]{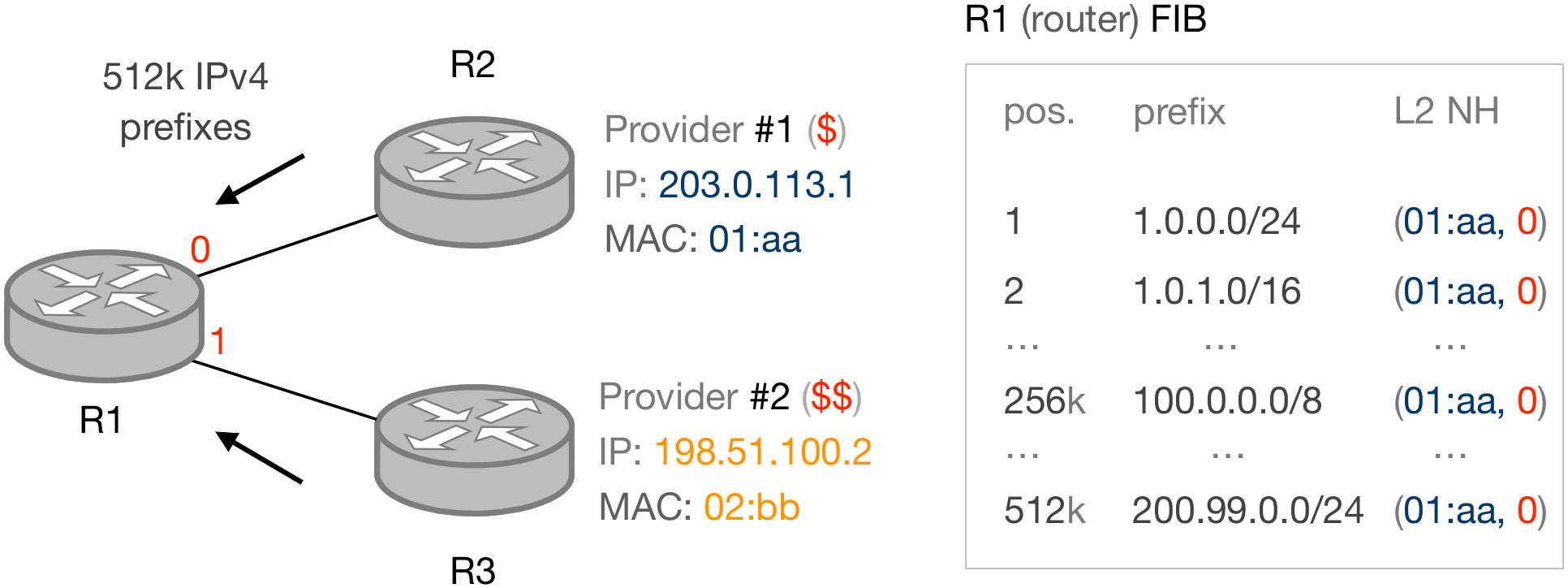}
 \caption{In a classical router, the Forwarding Information Base (FIB) is flat,
meaning each entry points to the actual physical L2 NH. Upon a failure of $R2$,
every single entry (512k) has to be updated to restore full connectivity, a
time-consuming operation.}
 \label{fig:fib_router}
\end{figure}

\myitem{Today's (slow) convergence}. The convergence time of traditional IP
routers is directly linked to the time it takes for the router to update its
hardware-based Forwarding Information Base (FIB) after it detects the failure.
To achieve fast lookup and limit memory cost, the FIB only contains the
information strictly necessary to forward packet. In the case of Ethernet
interface, each FIB entry maps an IP destination to the L2 NH address
(\emph{i.e.}, MAC address) of the chosen IP NH as well as the output
interface.

In most routers, the FIB is flat, meaning each FIB entry is mapped to a
different (but possibly identical in content) L2 NH entry. As an illustration,
consider the network depicted in Fig.~\ref{fig:fib_router}. $R1$ is an edge
router connected to the router of two providers, $R2$ and $R3$. Each of these provider routers advertise a full
Internet routing table composed of more than 512{,}000 IPv4
prefixes~\cite{cidr_report}. Also, as $R2$ is cheaper than $R3$, $R1$ is
configured to prefer $R2$ for all destinations. In such a case, each of the
512k FIB entries in $R1$ is associated to a distinct L2 NH entry which all
contain the physical MAC address of $R2$ ({\sffamily 00:aa}).

Upon the failure of a $R2$, every single entry of $R1$ FIB has to be updated
creating a significant downtime. Our measurements on a recent router (see
\S\ref{sec:evaluation}) shows that it actually takes \emph{several minutes} for
$R1$ to fully converge, during which traffic is lost. With the ever rising cost of downtime~\cite{cerin2013downtime} and as services increasingly
rely on high-availability, convergence of the order of minutes is simply not acceptable.

\begin{figure}[t]
 \centering
 \includegraphics[width=\columnwidth]{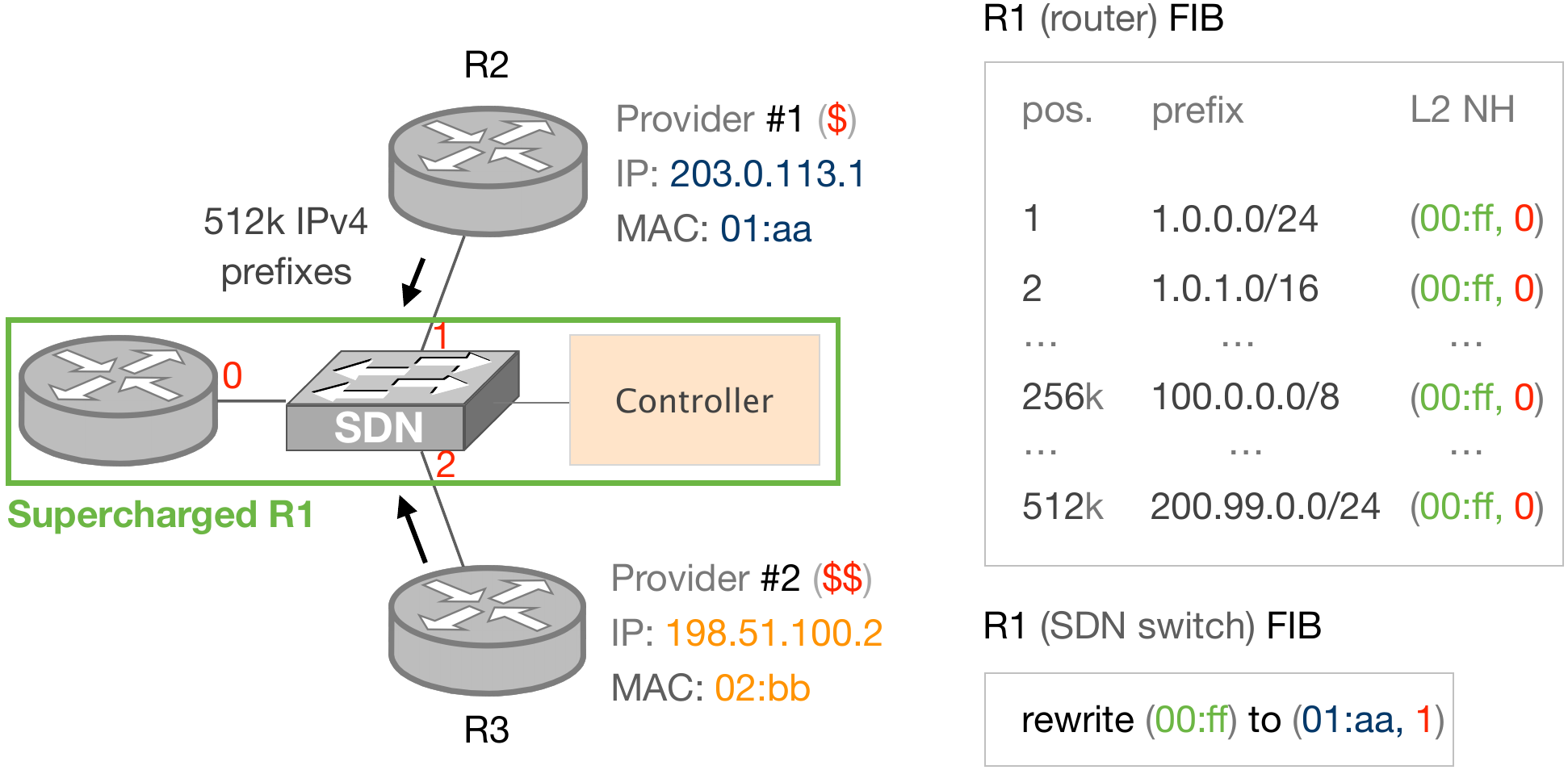}
 \caption{In a supercharged router, the combined FIB is hierarchical, each FIB
entry in the router points to a virtual L2 NH or pointer that is resolved in
the SDN switch. Upon failure of $R2$, only \emph{one entry}---the pointer
value---needs to update to restore full connectivity.\vspace{-0.5cm}}
 \label{fig:fib_supercharged}
\end{figure}

\myitem{Supercharging convergence.} Equipping routers with a
hierarchical FIB~\cite{FMBDVSBF11} is an obvious solution to
the convergence problem mentioned above. In a hierarchical FIB, each IP destination is mapped to a pointer that resolves to the actual L2 NH to be used. Upon failure of a L2 NH, only pointer values have to be updated. Since the number of L2 NH is several order of magnitude smaller than the number of FIB entries, convergence is greatly improved. Unfortunately, hierarchical FIB designs also means much more complex hardware, and therefore, more expensive routers.

Fig.~\ref{fig:fib_supercharged} illustrates how we can provide \emph{any}
router (here $R1$) with a hierarchical FIB, spanning two devices, by combining
it with a SDN switch. To provision forwarding entries in this hierarchical FIB,
we built a \emph{supercharged controller}. While the controller can rely on
(typically) OpenFlow to provision forwarding entries in a SDN switch,
dynamically provisioning specific forwarding entries in a router is trickier.
Our key insight is that the supercharged controller can use any routing
protocol spoken by the router as a provisioning interface. Indeed, FIB entries
in a router directs traffic to the L2 NH associated to the L3 NH learned via
the routing protocol. Our supercharged controller interposes itself between the
router and its peers (we explain how to make this reliable in
\S\ref{sec:implementation}), computes primary and backup NH for all IP
destinations, and provisions L2 NH ``pointers'' by setting the IP NH field
to a virtual L3 NH that gets resolved by the router into a L2 NH using
{\sffamily ARP}. Upon failure of $R2$ in Fig.~\ref{fig:fib_supercharged}, all the controller has to do to convergence is to modify the switch rule to ({\sffamily rewrite(00:ff) to (02:bb,2)}) in order to converge \emph{all traffic} to $R3$.

\myitem{Contributions.} We make the following contributions:
\begin{itemize}[leftmargin=*]
	\setlength{\itemsep}{0pt}
	\item \textbf{Supercharging router convergence:} We propose novel ways to
combine SDN and legacy networking equipment to improve convergence times (\S\ref{sec:supercharging}).
	\item \textbf{Implementation:} We describe a fully working prototype
implementation of a supercharger controller, combining OpenFlow/Floodlight and
ExaBGP (\S\ref{sec:implementation}). Our implementation is efficient, reliable,
and can be used to supercharge \emph{any} router.
	\item \textbf{Hardware-based Evaluation:} We supercharged a hardware router
(Cisco Nexus 7k) and thoroughly evaluated its performance
(\S\ref{sec:evaluation}). To ensure precise measurements, we developed a
FPGA-based traffic generator which detects traffic loss within
70$\mu$s. With respect to the normal router convergence under similar
conditions, the supercharged version converged systematically within 150ms, a
900$\times$ reduction!
\end{itemize}





\section{Supercharging convergence}
\label{sec:supercharging}

In this section, we describe how to supercharge the convergence of any existing
router using SDN equipment to build a hierarchical forwarding table. 

\myitem{Overview.} Since the number of destinations is much greater than the
number of neighbors, many destinations (IP prefixes) will share the same
primary and backup NH. We refer to the couple (primary NH, backup
NH) as \emph{backup-group}. For instance, in
Fig.~\ref{fig:fib_supercharged}, all 512k prefixes share $(R2,R3)$ as
backup-group. If $R2$ fails, all entries will be rewritten to
point to $R3$.

\begin{figure}
 \centering
 \includegraphics[width=.65\columnwidth]{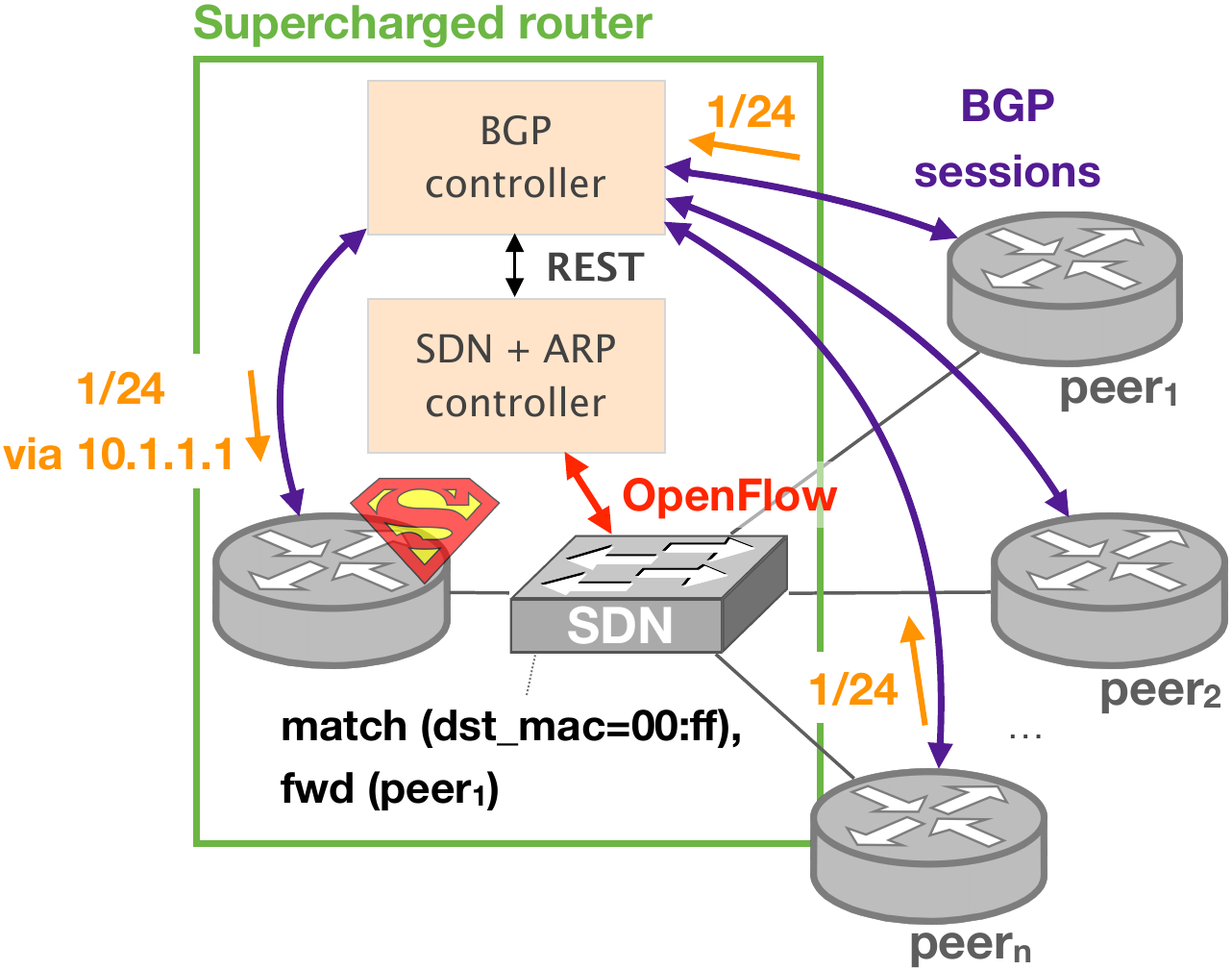}
 \caption{Supercharged router overview}
 \label{fig:overview}
 \vspace{-10px}
\end{figure}

In a supercharged router, we use the router to \emph{tag} the traffic according
to the backup-group it belongs to and use the switch to \emph{redirect} the
tagged traffic to the master or backup NH depending on its status. We use the
destination MAC address as the tag and provision it in the router using the virtual NH
field in routing announcements. Fig.~\ref{fig:overview} depicts the overall
architecture.

\myitem{Provisioning \emph{tagging entries} in the router's FIB.} To provision
entries in the router's FIB, a routing daemon is interposed between the router
and its peers. Its role is to compute the backup-groups for every IP
destination. For simplicity, we assume that BGP is used as routing protocol,
but other intra-domain routing protocols such as OSPF or IS-IS can also be used~\cite{fibbing_sigcomm_2015}. The
routing daemon assigns a Virtual IP NH (VNH) and a corresponding virtual MAC
(VMAC) address to each distinct backup-group and rewrites the routing NH in the
corresponding announcements that it directs to the supercharged router. In
Fig.~\ref{fig:overview}, the backup-group for 1.0.0.0/24 is $(peer_1,peer_n)$
and the corresponding (VNH, VMAC) is {\sffamily (10.1.1.1, 00:ff)}. Upon
reception of a route associated with a VNH, the router issues an ARP request to
resolve it to a MAC address. This ARP request is caught by the SDN controller
which replies with the corresponding VMAC address. After that, the supercharged
router will use the VMAC as the destination MAC for all the corresponding
traffic sent in the data-plane.

\begin{lstlisting}[caption={Online algorithm computes backup-group},label={alg:bck_group_computation}]
bck_groups = {}
routing_table = {}

def compute_backup_groups(bgp_upd):
 old = routing_table[bgp_upd.pfx]
 insert(routing_table, bgp_upd)
 new = routing_table[bgp_upd.pfx]
 
 if old:
  if not new:
   send_withdraw(bgp_upd.pfx)
  else:
    if new != old:
      if len(new) == 1:
       send(bgp_upd)
      else:
       if (new[0].nh, new[1].nh) != (old[0].nh, old[1].nh):
        if new[0].nh not in bck_groups:
         bck_groups[new[0].nh] = {}
        if new[1].nh not in bck_groups[new[0].nh]:
         bck_groups[new[0].nh][new[1].nh] = get_new_vnh_vmac()
        rewrite_nh(bgp_upd, bck_groups[new[0].nh][new[1].nh].nh)
        send(bgp_upd)
 else:
  send(bgp_upd)
\end{lstlisting}

\myitem{Computing backup-groups.} Listing~\ref{alg:bck_group_computation}
describes an online algorithm for computing the backup-group. In essence, the
algorithm maintains an ordered list of known NH for each IP prefix with the two
first elements identifying the backup-group. The algorithm sends a routing
update with a VNH whenever one of these elements change. Observe that the total
number of backup-groups depends on the number of peers $n$ the supercharged
router has. Taking into account all the neighbors of the supercharged router,
the total number of backup-groups is $\frac{n!}{(n - 2)!}$. For instance,
considering a router with 10 neighbors (a lot in practice), the number
of backup-groups is only 90. In this paper, we worked with backup-group of size 2, which can protect from any single link or node failure. Our algorithm in general though and can compute backup-groups of any size.

\myitem{Directing tagged traffic to the appropriate NH in the switch's FIB.}
The controller provisions dedicated flow entries to match on the VMAC
associated to each backup-group. By default, these rules direct the traffic to
the primary NH. Upon a node or a link failure, all the backup-group
entries for which the unreachable NH was the primary NH are rewritten to direct
the traffic to the backup NH instead. In the worst case, the number of flow
rewritings that has to be done is the number of peers of the supercharged
router, \emph{i.e.} a small constant value.
Listing~\ref{alg:data_plane_convergence} describes how the controller
determines what flow to install.

\begin{lstlisting}[caption={Data-plane convergence procedure },label={alg:data_plane_convergence}]
def data_plane_convergence(peer_down_id):
  for backup_nh in bck_groups[peer_down_id]:
    install_flow(
     match(dst_mac=bck_groups[peer_down_id][backup_nh].vmac),
     modify(dst_mac=get_mac(backup_nh)),
     fwd(output_port=get_port(backup_nh))
    )
\end{lstlisting}

\section{Implementation}
\label{sec:implementation}

We now briefly describe a reliable implementation of a supercharged controller. All our source code is available at {\small\url{https://github.com/nsg-ethz/supercharged_router}.}

\myitem{Controller.} We built our prototype atop ExaBGP~\cite{exabgp} as \emph{BGP controller}, FreeBFD~\cite{freebfd} as \emph{BFD daemon} (failure detection), and Floodlight~\cite{floodlight} as \emph{SDN controller}. 

ExaBGP enables us to establish BGP adjacencies and programmatically receive and send BGP routes over them. We extended ExaBGP with a complete implementation of the BGP Decision Process, the full algorithm to compute backup groups (see Listing~\ref{alg:bck_group_computation}) and the ability to rewrite BGP NH on-the-fly. FreeBFD provides a user-space implementation of the Bidirectional Forwarding Detection Protocol (BFD)~\cite{bfd_rfc}. We use it to speed up the discovery of peer failure. Upon a peer failure announcement produced by FreeBFD, ExaBGP uses the REST API provided by Floodlight to push the corresponding rewrite rules in the data-plane (see~\S\ref{sec:supercharging}). We also extended Floodlight with an {\sffamily ARP} resolver in order to reply to the {\sffamily ARP} queries generated by the router for resolving the virtual NH to the corresponding virtual MAC address.

\myitem{Reliability.} Any underlying SDN switch or any control-plane component of the supercharged controller can fail at any time. Since our goal is to enable fast convergence, our controller must be able to survive to any component failure to be of any use. Fortunately, reliability at both the data-plane and the control-plane is easily ensured.

At the data-plane level, reliability is obtained by using at least two
SDN-enabled switches connected to each supercharged router.
Observe that redundant SDN switches can be shared across multiple
supercharged routers that share physical connectivity, reducing the costs. At
the control-plane level, reliability is enforced by running at least 2 instances of the controller and connecting them to the corresponding supercharged router. Interestingly, no state needs to be synchronized across the backups as both backups will receive exactly the same input (BGP routes) and run the exact same deterministic algorithm and, hence, eventually compute the same outcome. The cost is the supercharged router to receive two copies of each route, and for the peers to configure an extra BGP session---slightly increasing the load in the control-plane. However, we note that control-plane memory is inexpensive (being classical DRAM) and routers maintain multiple BGP adjacencies already, for obvious redundancy reasons.

\section{Evaluation}
\label{sec:evaluation}

We now present a thorough evaluation of the convergence time of a recent
hardware router prior and after supercharging it using our prototype implementation. We then illustrate the scalability of our controller implementation using micro-benchmarks.

\begin{figure}[h]
 \centering
 \includegraphics[width=\columnwidth]{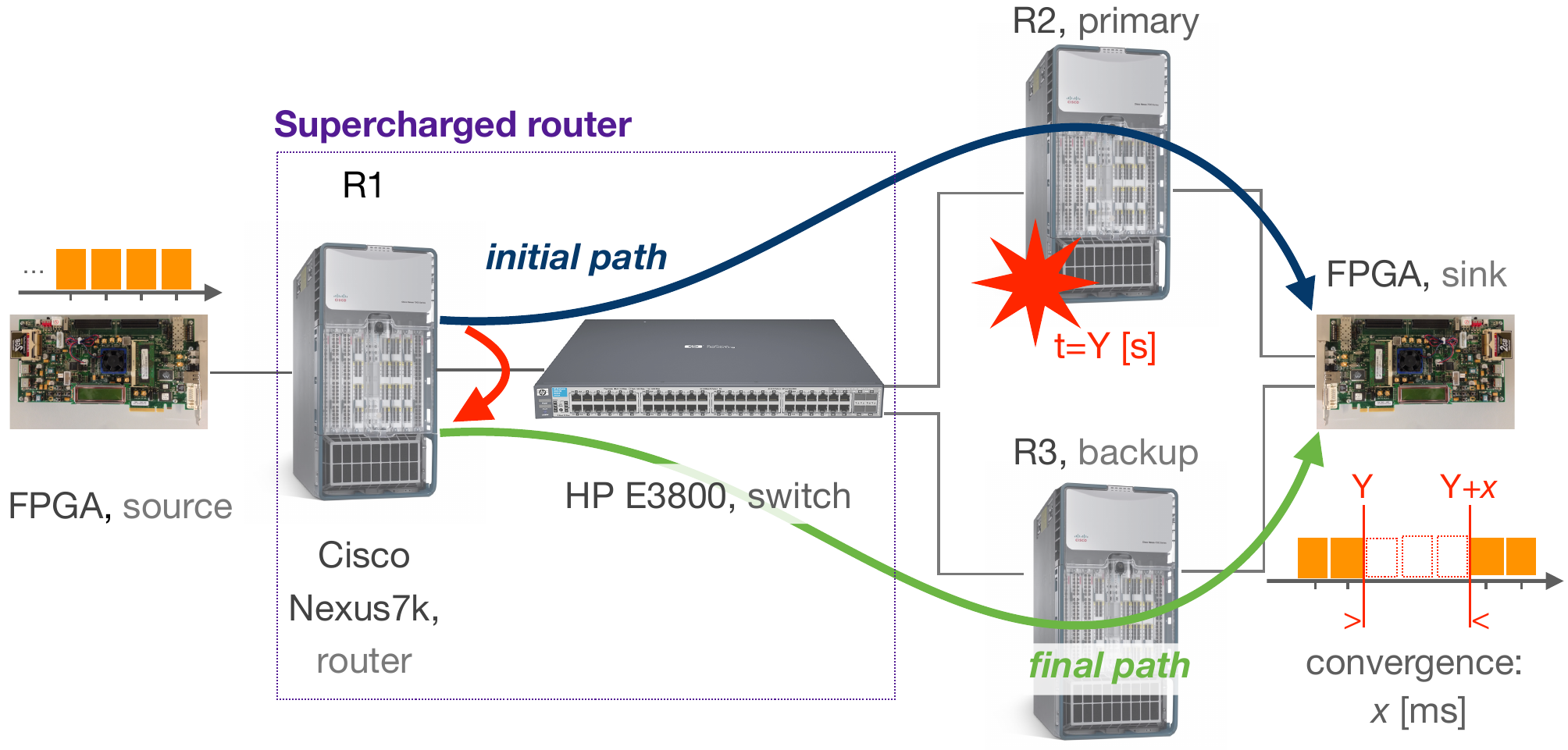}
 \caption{Overview of our HW-based convergence lab. $R2$ is rendered
inaccessible, causing $R1$ to switch to $R3$ for every single prefixes. At the
same time, we use FPGAs to precisely ($\mu$s resolution) measure the
convergence time. Ultimately, we compare the convergence time of the
supercharged R1 and the standalone R1. }
 \label{fig:lab_Setup}
\end{figure}

\myitem{Setup and methodology.} Our complete setup is depicted in Fig.\ref{fig:lab_Setup}. It consists of 3 routers Cisco Nexus 7k C7018 (running NX-OS v6.2, with no hierarchical FIB) interconnected through a HP E3800 J9575A Openflow-enabled switch. 

Using this setup, we measured the convergence time of $R1$ prior and after
supercharging it. To do so, we loaded $R2$ and $R3$ with an increasing number
of actual BGP routes collected from the RIPE RIS dataset~\cite{ripe:ris}. Both
$R2$ and $R3$ were loaded with the same feed to ensure that they both advertise
the same set of prefixes. In both cases (supercharged and not supercharged),
$R1$ was configured to prefer $R2$ for all the destinations. Once all routes
were advertised, we started to inject traffic at $R1$ using a FPGA-based
generator (see below). To compute a representative distribution of the
convergence time across different prefixes, we generated traffic towards 100 IP
addresses, randomly selected among the IP prefixes advertised by $R2$ and $R3$, and including the first and last prefix advertised. We configured $R2$ and $R3$ to send all receiving traffic to another FPGA-based
board, acting as sink. To ensure that the same detection time in both
experiments, we configured BFD on $R2$ on both experiments. We then
disconnected $R2$ from the switch, triggering the convergence process at $R1$;
subsequently, we measure the time until recovering full connectivity.

\myitem{Custom-built hardware-based traffic generator.} Since this project
deals with \emph{fast} convergence, we needed a way to accurately measure small
convergence time. Our choice rapidly went to hardware-based measurement, using
FPGA boards. Using the FPGAs, we were able to measure convergence time
\emph{with a precision of only 70 $\mu$s}. Such a precision would be impossible
to achieve using software-based measurements.

We measured the convergence time by monitoring the maximum inter-packet delays
seen by each flow between two FPGA boards: a source and a sink. For the FPGA
boards, we used a system-on-chip architecture with \emph{(i)} an embedded
MicroBlaze soft processor \emph{(ii)} an Ethernet MAC core, and (iii) either a
traffic generator (source) or traffic monitor (sink). The traffic monitor
matches the destination IP to a content-addressable memory (CAM) containing the
expected destination IPs, before it updates the corresponding maximum
inter-packet delay. We implemented both, source and sink, on Xilinx ML605
evaluation boards featuring a Virtex-6 XC6VLX240T-1FFG1156 FPGA.

We programmed the source FPGA to continuously send a stream of 64-byte UDP
packets to each of the 100 IPs over an 1G Ethernet connection. Doing so
generated a traffic load of about 725 MBit/s, which corresponds to about 1.4M
packets/s in total and 14K packets/s per flow.

\begin{figure}
 \centering
 \includegraphics[width=1\columnwidth]{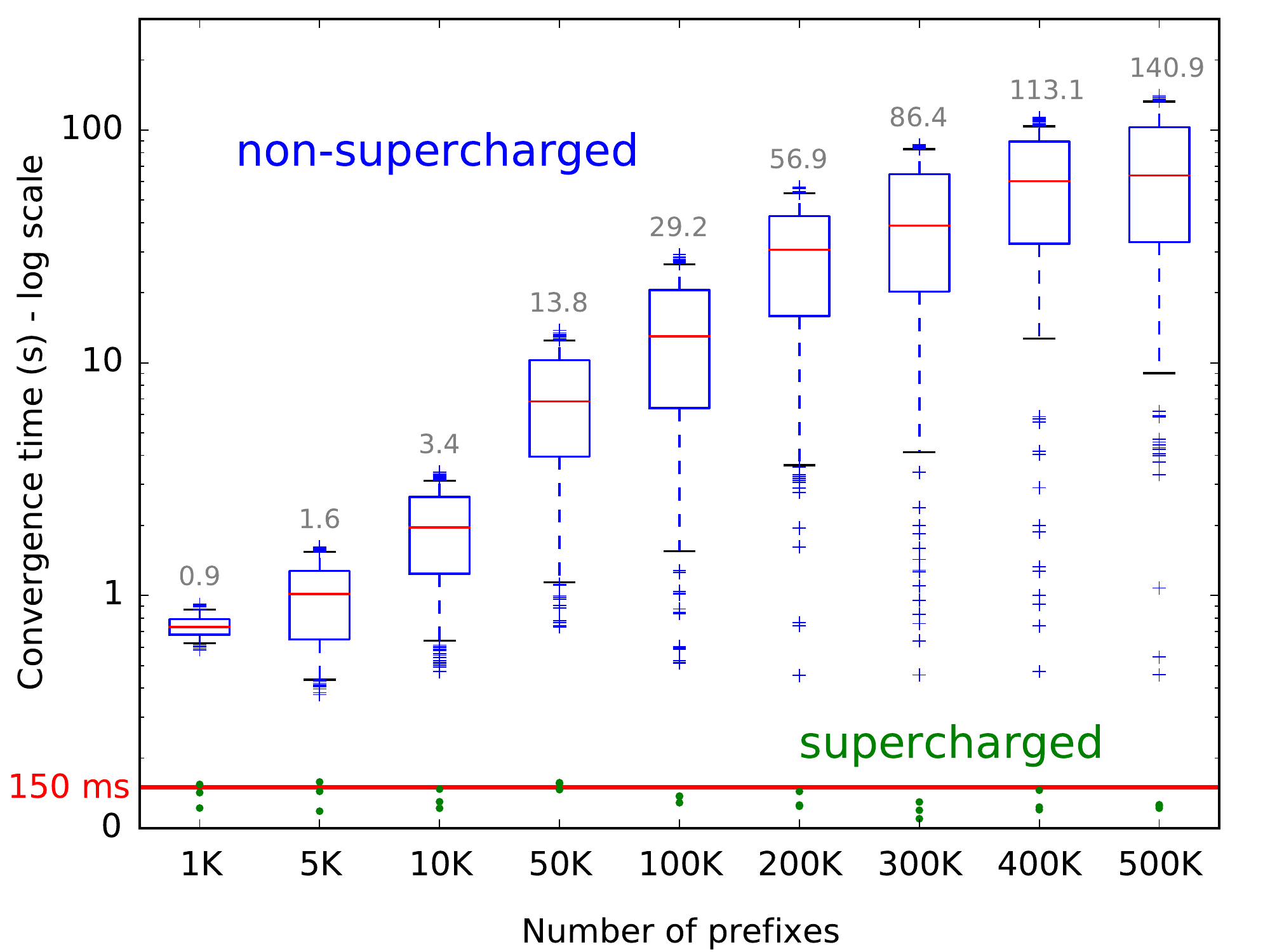}
 \caption{With respect to the normal convergence time which increases linearly with the number of prefixes, our supercharged router \emph{systematically converged within 150ms}. In contrast, the non-supercharged router took more than 2 minutes to converge in the worst-case.}
 \label{fig:convergence_time}
 \vspace{-10px}
\end{figure}

\myitem{The non-supercharged R1 took {\raise.17ex\hbox{$\scriptstyle\sim$}}2.5min to converge in the worst-case.} Using the methodology above, we measured the convergence time of the
router prior and after the supercharging process for an increasing number of
prefixes (from 1k to 500k). We repeated the experiment 3 times per number of
advertised prefixes. Since for each experiment, we measured the convergence of 100 prefixes, we ended up with 300 statistically representative data points per measurement. Fig.~\ref{fig:convergence_time} depicts the distribution of the convergence time using box-plots; both the non-supercharged and
supercharged routers are displayed. Each box shows the inter-quartile range of the convergence time; the line in the box depicts the median value; and the whiskers show 5th and 95th percentiles. The numbers on top are the maximal convergence time recorded.

For the non-supercharged R1, we can see that the convergence time is roughly
linear\footnote{The linearity of convergence time is not well reflected in Fig.~\ref{fig:convergence_time} because of the non-uniform scaling of the $x$-axis.} in the number of prefixes in the FIB. This is because FIB entries are updated one-by-one; while the first FIB entry is
updated immediately, irregardless of the total number of prefixes, the last
entry updated must wait for all the preceding FIB entries to be updated. This
worst-case highlights undesirability of the non-supercharged approach: as the
FIB grows, so does the convergence time. Here, we see that $R1$ took close than
2.5min to converge when loaded with 512k.

\myitem{The supercharged R1 systematically converged \emph{within 150ms}, for
all prefixes.} Thanks to its hierarchical FIB design, the supercharged R1's
convergence time was constant---irrespective of the number of prefixes. This is
illustrated in Fig.~\ref{fig:convergence_time} by a almost horizontal line
around 150ms. With respect to the above worst-case, this constitutes a
900$\times$ improvement factor. Interestingly, the worst-case convergence time
of a supercharged router is still more than two times faster that the best-case
convergence time of its standalone counterpart. Indeed, in the best case, it took 375 ms for the standalone R1 to update the first FIB entry.

\myitem{The supercharged controller processed each BGP update under 125ms.}
While supercharging router drastically improves its data-plane convergence
time, it slightly increases its control-plane convergence time due to the need
to re-compute the backup-group upon every BGP announcement and, potentially,
update the virtual NH. To quantify this overhead, we measured the time our
unoptimized, python-based BGP controller took to process two times 500K updates
from two different peers. 
In the worst-case, processing an update took 0.8s but the 99th percentile was only 125ms. We argue that this is a reasonable price to pay for improving the
convergence by several orders of magnitude.

\section{Related Work}
\label{sec:related_work}

\myitem{Routing.} The problem of minimizing down time during convergence has
been well studied in the domain of distributed routing
protocols~\cite{SBPFFB13, FB07, FFEB05, FMBDVSBF11}. Among all these works, BGP
Prefix Independent Convergence (PIC)~\cite{FMBDVSBF11} is certainly the most
relevant. PIC introduces the idea of using a hierarchical FIB design in
order to speed-up router convergence upon peering link failure. In essence, our supercharged router replicates the functionality of PIC but on \emph{any}
routers (even old ones), without requiring expensive line-cards update.

\myitem{SDN.} FatTire~\cite{reitblatt2013fattire} is a domain-specific language
which aims at simplifying the design of fault-tolerant network programs that
can quickly converge by leveraging fast-failover mechanisms provided in recent
versions of OpenFlow~\cite{openflow1.4}. While FatTire targets fully-deployed
OpenFlow networks, we show that we can already speed up the convergence of
existing network with a single SDN switch. In~\cite{gamperli2014evaluating},
Gamperli \emph{et al.} evaluated the effect of centralization on BGP
convergence. They showed that the convergence time decreases as more and more
of the network-wide decision get centralized. Supercharging routers is a direct
complement to their work. Once enough routers have been supercharged, one can
use~\cite{gamperli2014evaluating} at the network-level to speed-up convergence
even more. Just as a supercharged router, SDX~\cite{sdx_sigcomm2014} is also an
example of how routing and SDN can coexist in a symbiotic way, providing each
other benefit. While SDX showed how router can boost SDN equipment performance,
we show how SDN equipment can boost router performance. Also, our technique can immediately be applied to the SDX environment in order to boost the convergence time upon the failure of an IXP participant equipment.

\myitem{Incremental SDN deployment.} RouteFlow~\cite{routeflow} and
Panopticon~\cite{Panopticon-atc14} proposed techniques to incrementally deploy
SDN equipments in existing networks with the aim of reaping early benefits.
RouteFlow enables operator to build fully-fledged IP router out of a SDN
switch, while Panopticon enables to steer traffic away from a L2 domain to SDN
equipment where it could be processed. In contrast to supercharging routers,
none of them improve the performance of existing equipment.
In~\cite{agarwal2013traffic}, Agarwal \emph{et al.} proposed a way to improve
the Traffic Engineering (TE) performance of existing networks even in partial
deployment of SDN capability, highlighting another aspect of the network that can be ``supercharged'' using SDN devices.

\section{Conclusions}
\label{sec:conclusion}

We boost the convergence time of legacy routers by combining them with SDN
equipment in a novel way, essentially building a hierarchical forwarding table
spanning across devices. Through thorough evaluations on real hardware, we demonstrated significant gains with convergence time reduced by up to 900$\times$. 

We believe this paper opens up many interesting future directions for
integrating legacy routing and SDN devices in a more ``symbiotic way''. By
juxtaposing the agility of the SDN with the tried-and-true routers prevalent in
the industry today, we take the best of both worlds and take the first steps
towards electrifying modern day networks through supercharged networking
devices.

\bibliographystyle{IEEEtran}
\bibliography{main}

\end{document}